\documentclass[conference]{IEEEtran}
\IEEEoverridecommandlockouts
\usepackage{cite}
\usepackage{amsmath,amssymb,amsfonts}
\usepackage{graphicx}
\usepackage{textcomp}
\usepackage{xcolor}
\usepackage{cellspace} 
\usepackage{makecell} 
\setcellgapes{1pt}
\usepackage{booktabs} 
\usepackage{multirow}
\usepackage{amsthm}
\newtheorem{definition}{Definition}
\usepackage{algorithm}
\usepackage{algorithmicx}
\usepackage[noend]{algpseudocode}

\algrenewcommand\alglinenumber[1]{\tiny #1:}
\def\BibTeX{{\rm B\kern-.05em{\sc i\kern-.025em b}\kern-.08em
    T\kern-.1667em\lower.7ex\hbox{E}\kern-.125emX}}

\begin{document}

\title{ReIGNN: State Register Identification Using Graph Neural Networks for Circuit Reverse Engineering}
% I think the title ReIGNN: Register Identification Using Graph Neural Networks for Circuit Reverse Engineering is not sounding correct. It means we are identifying registers which is not true. We are separating them as State and non-state registers.

\author{\IEEEauthorblockN{Subhajit Dutta Chowdhury, Kaixin Yang, Pierluigi Nuzzo}
\IEEEauthorblockA{Ming Hsieh Department of Electrical and Computer Engineering, University of Southern California, Los Angeles, CA \\
\{duttacho, kaixinya, nuzzo\}@usc.edu}
}

\newcommand{\pierluigi}[1]{\normalsize{\color{magenta}(PN:\ #1)}}
\newcommand{\Subhajit}[1]{\normalsize{\color{pink}(SDC:\ #1)}}
\newcommand{\kaixin}[1]{\normalsize{\color{orange}(KY:\ #1)}}

\maketitle

\begin{abstract}
Reverse engineering an integrated circuit netlist is a powerful tool to help detect malicious logic and counteract design piracy. A critical challenge in this domain is the correct classification of data-path and control-logic registers in a design. 
% Existing techniques based on the detection of similarities between the topologies of the  registers' fan-in circuits tend to heavily rely on fine-tuning of a set of configuration parameters for each circuit to achieve high classification accuracy. 
%%% I think we need to modify a bit since we need to say that these techniques are for identifying the control logic registers. Right now it is broader. I am writing the modified line below. It is a very small modification.
%A critical challenge in control logic reverse engineering is the correct identification of the control/state registers in a design. Existing techniques based on the detection of similarities between the topologies of the  registers' fan-in logic circuit tend to heavily rely on fine-tuning of a set of configuration parameters for each circuit to achieve high identification accuracy.
% \pierluigi{In this paper, we present ReIGNN, a register classification methodology, that combines graph neural networks (GNN) with structural analysis to achieve  higher classification accuracy and generalization properties.}
% In this paper, 
We present ReIGNN, a novel learning-based register classification methodology that combines graph neural networks (GNNs) with structural analysis to classify the registers in a circuit with high accuracy and generalize well across different designs.
% REIGNN represents the design netlist as its intrinsic data structure, a graph, and utilizes
% \pierluigi{GNNs leverage training and inference (processing) on graph to efficiently/effectively learn structural properties of circuits, including properties of their nodes and the corresponding neighborhood.} \pierluigi{Explain why GNNs might be promising or powerful for learning structural information from IC netlists.}
% \Subhajit{
GNNs are particularly effective in processing circuit netlists in terms of graphs and leveraging properties of the nodes and their neighborhoods to learn to efficiently discriminate between different types of nodes.
% adeptly perform different graph-related tasks like node classification. They 
% information from the inherent graph structure 
% GNNs are a good fit for our target problem since the intrinsic data structure of a circuit is a graph. The key here is to leverage well trained GNN to classify the registers.
% }  
% Finally these node embeddings are fed to the downstream machine learning pipeline to perform the register classification task. 
% \textcolor{blue}{
Structural analysis can further rectify any registers misclassified as state registers by the GNN by analyzing strongly connected components in the netlist graph.
% }
% 
Numerical results on a set of benchmarks show that ReIGNN can achieve, on average, 96.5\% balanced accuracy and 97.7\% sensitivity across different designs.  
\end{abstract}


\section{Introduction}

Integrated circuits (ICs) are often considered the root of trust of modern computing systems. However, the  globalization of the IC supply chain\cite{DARPA} and the reliance on third-party resources, including third-party intellectual property (IP) cores and fabrication foundries, 
% and commercial off-the-shelf ICs 
have brought into question IC protection against threats like IP piracy,  counterfeiting~\cite{quadir2016survey,guin2014counterfeit}, and hardware Trojan (HT) insertion~\cite{tehranipoor2010survey}. 
% Devising techniques that can ensure the trustworthiness of the fabricated chips becomes a necessity. 
Reverse engineering is a promising tool to 
% counteract design piracy \pierluigi{how?} and 
restore trust in the IC supply chain by enabling the identification of malicious modifications in a design. 
% and to discover possible patent violations 

Hardware reverse engineering consists in extracting the  high-level functionality of a design through multiple automated or manual analysis steps. The first step is to obtain the gate-level netlist from a fabricated chip through delayering, imaging, and post processing~\cite{torrance2009state, quadir2016survey,baehr2020machine}. Once the netlist is available, gate-level netlist reverse engineering techniques are employed to infer the functionality of the netlist. Netlist reverse engineering consists of two main tasks, namely, data-path reverse engineering and control-logic reverse engineering. The control logic in a digital circuit is in charge of generating signals to control the data path. Data-path blocks are more suitable for algorithmic analysis, due to their regularity and high degree of replication, using techniques like template matching~\cite{gascon2014template}, behavioral matching of an unknown sub-circuit against a library of abstract components~\cite{li2012reverse} or machine learning (ML)-based circuit recognition~\cite{fayyazi2019deep}. On the other hand, the control logic of a design is generally designed for a specific functionality, which makes it difficult to apply the techniques mentioned above.  
% for sequential control logic making their reverse engineering a hard problem. 
Reverse engineering the control logic involves the  identification of the state registers followed by a finite state machine (FSM) extraction step, which recovers the state transition graph (STG)~\cite{shi2010highly, mcelvain2001methods,meade2016netlist,fyrbiak2018difficulty} of the circuit FSM. Correct identification of the state registers is, therefore, instrumental in the extraction of a correct FSM.
% The extraction algorithm can provide a correct FSM only if the state registers have been correctly identified. 

Many techniques have been proposed to identify the state registers~\cite{meade2016gate, brunner2019improving, geist2020relic} in a design. Techniques like RELIC~\cite{meade2016gate} and fastRELIC~\cite{brunner2019improving} are based on the detection of similarities between the topologies of the registers’ fan-in circuits. However, they tend to depend on fine-tuning of a set of configuration parameters for each circuit to achieve high accuracy.
%since there exists no method that can provide the optimal parameter settings for different design netlists.
When little or no information is available about a design netlist, it becomes challenging to converge on the parameter values which can lead to the best result. Moreover, the complexity of these techniques varies polynomially with the number of registers in a design, i.e., the higher the number of registers, the higher the classification time. %\pierluigi{Are we independent of the size?} \Subhajit{In their case the complexity is polynomial while in our case once the model is trained, the inference occurs instantly for all the registers in the design}.

Inspired by the success of state-of-the-art deep learning (DL) methods in solving different challenging problems across different fields, we present ReIGNN (Register Identification Using Graph Neural Networks), a novel learning-based technique to classify the state and data registers in a netlist. ReIGNN combines GNNs with structural analysis to achieve high accuracy and generalization properties across different designs. GNNs are deep learning models that operate directly on graphs and adeptly perform different graph-related learning and inference tasks like node classification by leveraging the graph structure, the properties of the nodes, and the characteristics of their neighborhoods. 
% to learn a representation for each node that help discriminate between different types of nodes. 
% GNNs are an apt choice for our problem statement since the intrinsic data structure of a circuit netlist is a graph where 
By mapping the circuit gates and registers to graph nodes and their connections to graph edges, the register classification task directly translates into a node classification problem. In ReIGNN, we automatically abstract the circuit netlist as a graph and associate each graph node with a feature vector that captures information about its functionality and connectivity. We then train the GNN that processes these graphs to  discriminate between state and data registers in the netlist. The trained GNNs can then be used to classify the registers of new netlists. % \textcolor{blue}{
Finally, we perform structural analysis of the netlist and, specifically, identification of strongly connected components that include registers in the graph, to further improve the accuracy of the GNN and rectify misclassifications.
In summary, this paper makes the following  contributions:
\begin{itemize}
  \item 
  % \textcolor{blue}{
  The identification and efficient extraction of a set of features for each gate and register in a netlist to capture their functionality as well as connectivity properties.
  % }
  \item ReIGNN, a novel learning-based methodology which combines GNNs with structural analysis to classify the registers in a circuit. To the best of our knowledge, this is the first GNN-based circuit register classification method.  
\end{itemize}

\noindent Empirical results on a dataset containing standard circuit designs show that ReIGNN can achieve, on average, 96.5\% balanced accuracy and 97.7\% sensitivity  across the different designs.  
% \kaixin{to complete after having experimental results}. 

The remainder of this paper is structured as follows. In Section~\ref{sec:related_work} we discuss the related work. Section~\ref{sec:prelim} provides background notions at the basis of our technique. We describe the learning-based register classification methodology in Section~\ref{sec:reginn} and its detailed evaluation in Section~\ref{sec:eval}. Finally, we conclude in Section~\ref{sec:conclusions}. 

\section {Related Work}\label{sec:related_work}
%\subsubsection {Register Identification Techniques} \label{2.2}

\begin{figure}[t]
  \centering
  \includegraphics[width=\columnwidth]{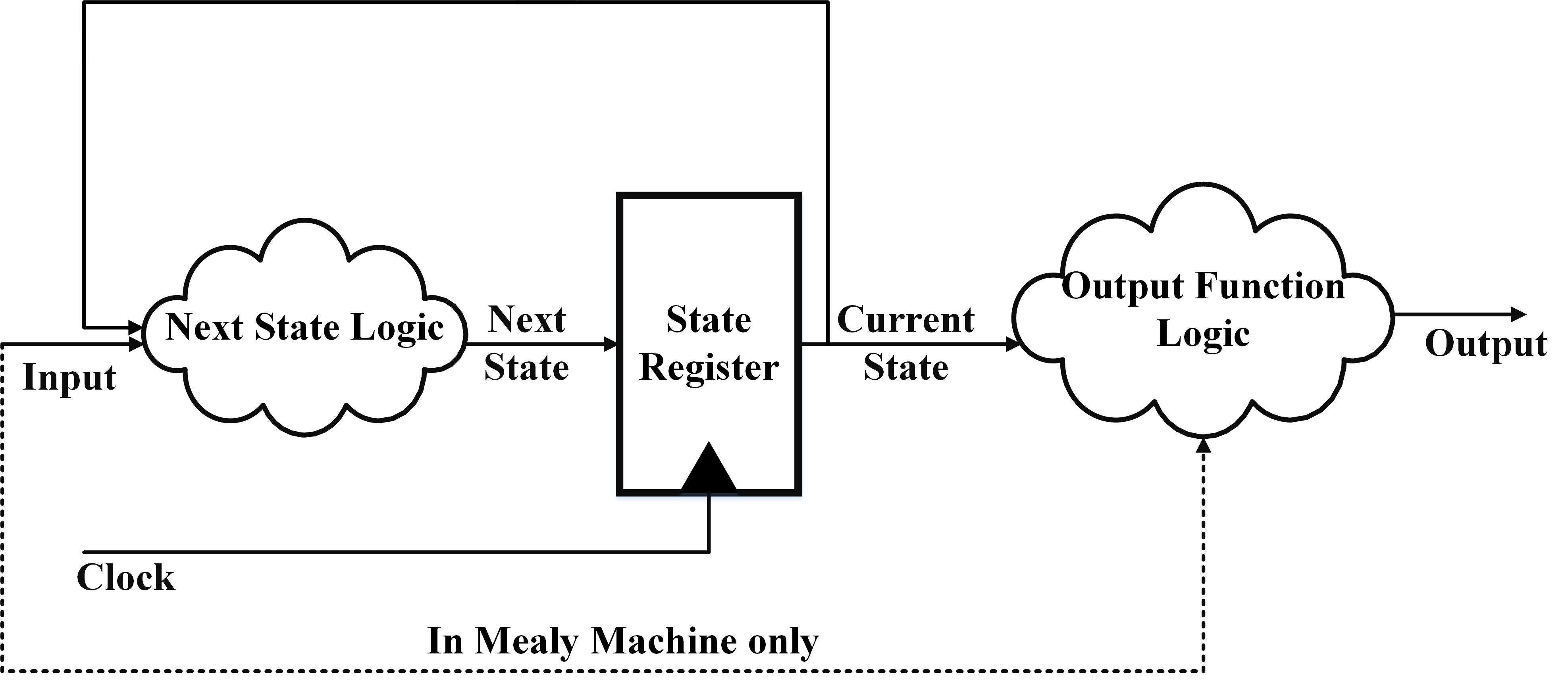}
  \caption{Block diagram of an FSM. In a Moore machine, the output function logic depends on the current state only. In a Mealy machine, the output is a function of both the current state and the input (dotted line).}
  \label{fig:fsm}
  \vspace{-4mm}
\end{figure}

Reverse engineering the control logic of a circuit aims at extracting an FSM from the gate-level netlist~\cite{brunner2019improving}. The first and often most critical step in FSM extraction is the correct classification of the state and data registers. Many techniques have been proposed over the years to perform this task. A register can be classified as a state register if it has a feedback path~\cite{mcelvain2001methods}, based on the observation that the state registers in an FSM always have a logic path from their output to the input, as shown in Fig.~\ref{fig:fsm}. However, the accuracy of a technique exclusively based on this rule tends to be poor since it neglects that many data-path designs employ data registers including feedback paths. Shi~et~al.~\cite{shi2010highly} improve the classification accuracy by eliminating all the candidate state registers that do not affect a control signal like the select line of a multiplexer or the enable input of a register. 
%Making assumption that the control signals can be identified in a design does not hold true besides after logic synthesis, these signals may have a different connection. 
More recently, the reset signal, control behavior, and feedback path information associated with the registers have been used together with strongly connected component (SCC) analysis to improve on the results of previous methods~\cite{fyrbiak2018difficulty}. While ReIGNN leverages graph algorithms and SCC, it differs from previous techniques in that it also uses the structural and functional properties of each element of the netlist.

Another category of reverse engineering techniques compare the registers' fan-in circuits either topologically or functionally to determine the type of registers, by exploiting the replication of circuit structures occurring naturally in the implementation of multi-bit data-path designs. RELIC~\cite{meade2016gate} performs a similarity test between all pairs of registers' fan-in circuits to identify potential state registers. Given two registers, it computes a ``similarity'' score between $0$ and $1$. 
% which denotes how much similar the two registers' fan-in logic circuit are topologically. 
% RELIC compares every register pair in the design. 
Intuitively, data-path registers may be clustered into groups that implement the same functionality, hence they have similar fan-in circuit structure. On the other hand, state registers do not generally have similar fan-in circuit structures because the FSM state transitions tend to depend on unique logic conditions. Different state registers tend to relate to different transition logic blocks. The registers are then classified based on pre-defined threshold values and the similarity scores. 
fastRELIC~\cite{brunner2019improving} improves the performance of RELIC by reducing the number of register pair comparisons and by considering not only the similarity between the registers' original fan-in circuits, but also the similarity between the original and the inverse fan-in circuits, obtained by replacing the AND and OR gates with OR and AND gates, respectively, and inverting the inputs~\cite{meade2018old} of the original fan-in circuits.
%\pierluigi{what is this?} 
%It is functionally very similar to RELIC; it addition to calculating the similarity score between the registers' fan-in circuits, it also considers the similarity score between the original and the inverse fan-in circuits. 
A greedy register clustering algorithm is used to group the registers based on the similarity of the structure of their fan-in circuits. Instead of comparing each register with every other register in the design, the comparison is performed with only one member of a  cluster. However, the choice of the member is based on heuristics, and the worst-case complexity is the same as RELIC. 
% ~\cite{brunner2019improving}. 
%%The complexity analysis in~\cite{brunner2019improving} shows that in the best case, fastRELIC requires linear number of comparisons but in the worst case, it requires the same number of comparisons as that of RELIC. 

Subramanyan et al.~\cite{subramanyan2013reverse, subramanyan2013reversejournal} compare the  registers' fan-in logic functions rather than their structure. 
% ally, i.e., in terms of the Boolean function. 
Cut-based Boolean function matching is used to find replicated bit-slices in the netlist so that all the data registers can be identified and aggregated into words. A bit-slice is a Boolean function with one output and a small number of inputs. 
%to aggregate them into multibit components i.e., data registers.
%storing the output of adders, subtractors, or multipliers. Here, Boolean matching is used to find data registers and aggregate them into words. 
RELIC-FUN~\cite{geist2020relic} uses the same algorithmic approach 
% as~\cite{subramanyan2013reverse, subramanyan2013reversejournal} 
but focuses on finding registers whose fan-in logic implements different functions. 
% \textcolor{blue}{
Unlike these techniques~\cite{meade2016gate, brunner2019improving, geist2020relic}, ReIGNN does not require pairwise comparisons of the structure or function of the registers' fan-in circuits or pre-calibration of threshold values for similarity scores, since the registers of an unknown circuit are directly classified by the trained GNN.
%}

DL has been recently used for different tasks in the context of hardware security, e.g., to detect HTs~\cite{tehranipoor2010survey, huang2020survey, yasaei2021gnn4tj},  attack logic locking~\cite{hu2021risk, chowdhury2021enhancing, sisejkovic2021challenging, chakraborty2018sail}, or identify the functionality of data-path building blocks~\cite{fayyazi2019deep, baehr2020machine}. GNNs have also been used for attacking logic locking~\cite{alrahis2021gnnunlock}. Our work is different, since it uses DL models, and  specifically GNNs, for classifying the registers in a netlist, which is the first step toward reverse engineering the circuit control logic.    
%\pierluigi{Too many details for each technique. Unclear how we compare with all the related work.}
%Hence, the use case of the algorithm differs in the two cases, where in one case, the algorithm is used for data word aggregation whereas in the other case, it is used to isolate the state registers. 

\section{Preliminaries}\label{sec:prelim}

% \textcolor{blue}{
This section outlines some background concepts that will be used throughout the paper.
% } 

\subsection {FSM Model} \label{sec:prelim_fsm}

% While formulating the problem statement for control logic reverse engineering, it is important we make some assumptions about the control logic model. 
We model the control logic of a circuit as an FSM, a computational model characterized by a finite number of states. Depending on the input and the current state, the FSM transitions to the next state at each reaction and generates output control signals~\cite{fyrbiak2018difficulty}. 

% Formally, an FSM can be defined as: \par

\begin{definition}[FSM]
An FSM is a 6-tuple \( (S, I, s_0, \delta, O, \omega) \) where \(S\) is the finite set of states, \(I\) is the finite set of inputs, \(O\) is the finite set of outputs, \(s_0 \in S\) is the initial state,  \(\delta: S \times I \rightarrow S\) is the state transition function which determines the next state depending on the input and the current state, and \(\omega\) is the output function logic.
\end{definition}
%\pierluigi{Use latex environments for definitions.} 
Figure~\ref{fig:fsm} shows the block diagram representation of an FSM which consists of: (a) the next state logic computing \(\delta\), (b) a state register that stores the current state \(S\), and (c) the output function logic that computes \(\omega\). 
%\pierluigi{The figure should appear after it is named or in the same page.}
% FSMs can be broadly classified into 2 groups - Moore and Mealy machines based on their output function logic \(\omega\). In case of Moore Machine, the output function logic depends only on the current state (\(\omega: S \rightarrow O \)) whereas in Mealy machines the output depends on both the current state and input (\(\omega: S \times I \rightarrow O \)). 
All the state registers in an FSM are connected to the same \(clock\) signal, which is used to trigger the FSM reaction. The states can be encoded using different styles  to meet different design goals. 
% like minimizing the power consumption and area while maximizing the speed. 
% Most commonly used state encoding styles are 1-hot encoding and binary encoding. 
For example, when using one-hot encoding,  \(|S|\)-bit registers are required to represent the state, $|S|$ being the cardinality of the set $S$. The number of required registers is higher than with binary encoding, which only uses \(\lceil \log_2 |S| \rceil\)-bit registers~\cite{hennessy2011computer}. 
% to represent a state of the FSM.
%the complexity of the next state logic tends to be lower when adopting one-hot encoding. %\pierluigi{Reference here?}
% simple. In  Binary encoding uses minimum number of registers to represent the state of an FSM resulting in more complex combinational logic to design the next state logic. 
%Since, in most of the designs the control logic is implemented as an FSM, so in this work, without loss of generality, we model the control logic as an FSM. 
In this paper, we denote a memory element that stores one bit of the encoded state of an FSM as \emph{state register}. All other memory elements in the design are denoted as \emph{data registers}. 
% \Subhajit{This statement clarifies how we define state and data registers.} 

\subsection{RELIC} \label{sec:prelim_relic}

We detail RELIC~\cite{meade2016gate}, which will be used as a reference method to compare with ReIGNN. RELIC pre-processes a gate-level netlist such that it contains only AND, OR, and INV gates, to ensure that similar logic functions are represented by similar circuit structures~\cite{meade2016gate}. 
% Once the netlist is pre-processed, now 
A topological similarity test is then performed on the pre-processed netlists to calculate the similarity score between every register pair, as denoted by the \textsc{SimilarityScore} function in Algorithm~\ref{alg:relic}. The calculated similarity score lies between $0$ (no similarity) and $1$  
% where a score of 0 means no similarity while a score of 1 denotes 
(identical fan-in circuit structure). 
% \Subhajit{
RELIC first creates a bipartite graph with the fan-in nodes of the registers. The fan-in nodes are then compared recursively and, if their similarity score is above $T1$, an edge is connected to the bipartite graph. To avoid a large number of recursive calls to the \textsc{SimilarityScore} function, the analysis of the fan-in circuits is limited to a maximum depth \emph{d}. A maximum matching algorithm is then executed on the bipartite graph to generate the similarity score between two registers. If this similarity score is above $T2$, then the similarity count of both registers is incremented by one. Finally if the similarity count of a register is above $T3$, then it is classified as a data register. RELIC requires the user to define these three threshold parameters, $T1, T2$, and $T3$, together with the depth of the fan-in circuits.
% }
%\pierluigi{What are the threshold for? Can you provide a bit more details to better understand the algorithm?} \kaixin{These three threshold parameters provide numerical standards for the definition of similar children, similar registers and state registers.}

\begin{algorithm} 
    \scriptsize
	\caption{RELIC} 
	\label{alg:relic} 
	\begin{algorithmic}[1]
	    \Require{\\ $graph$: the graph generated from the netlist \\ $registers$: the list of registers in the netlist \\ $T1, T2, T3$: three threshold values selected by the user to tune the result of RELIC \\ $d$: depth of the registers' fan-in logic circuits}
	    \State
	    \Function{SimilarityScore}{$i, j, d, T1$}
		    \State $max\_children \gets \textsc{max}(i.num\_children, j.num\_children)$
		    \State $min\_children \gets \textsc{min}(i.num\_children, j.num\_children)$	        
		    \If{$i.type \not= j.type$}
		        \State \Return{0}
		    \EndIf
		    \If{$d == 0$}
		        \State \Return{$min\_children/max\_children$}
		    \EndIf
	        \State insert all children nodes of i and j into a new bipartite graph G
		    \For{$a \in i.children$}
	            \For{$b \in j.children$}
	                \State $score \gets \textsc{SimilarityScore}(a, b, d - 1, T1)$
	                \If{$score > T1$}
	                    \State add edge between a and b in G
	                \EndIf
	            \EndFor
	        \EndFor
            \State \Return{$\textsc{maxmatching}(G)/max\_children$}
	    \EndFunction
	    \State
		\Function{ClassifyRegister}{$graph, registers, d, T1, T2, T3$}
		    \State $n \gets$ length of registers
		    \State $sim\_score \in \mathbb{R}^{n \times n}$
		    \State $pair\_cnt \in \mathbb{Z}^n$
		    \State $register\_type \in$ \{control, data\}$^n$
		    \For{$i \in registers$}
		        \For{$j \in registers$}
		            \If{$i == j$}
		                \State $sim\_score[i][j] = 1$
		            \Else
		                \State $sim\_score[i][j] = \textsc{SimilarityScore}(i, j, d, T1)$
		            \EndIf
                    \If{$sim\_score[i][j] > T2$}
		                \State $pair\_cnt[i] += 1$
		                \State $pair\_cnt[j] += 1$
		            \EndIf
		        \EndFor
		    \EndFor
		    \For{$reg \in registers$}
		        \If{$pair\_cnt[reg] \leq T3$}
		            \State $register\_type[reg] = $ control
		        \Else
		            \State $register\_type[reg] = $ data
		        \EndIf
		    \EndFor
		\EndFunction
	\end{algorithmic}
\end{algorithm}
\vspace{-3mm}

\subsection {Graph Neural Networks} \label{sec:prelim_gnn}

End-to-end deep learning paradigms like convolutional neural networks (CNNs)~\cite{lecun1995convolutional} and recurrent neural networks (RNNs)~\cite{hochreiter1997long} have shown revolutionary progress in different tasks like image classification~\cite{girshick2014rich, he2016deep}, object detection~\cite{redmon2016you}, natural language processing~\cite{young2018recent}, and speech recognition~\cite{hinton2012deep}. CNNs are appropriate for processing grid-structured inputs (e.g., images), while RNNs are better used over sequences (e.g., text). Both CNNs and RNNs are less suited to process  graph-structured data, possibly exhibiting  heterogeneous topologies, where the number of nodes may vary, and different nodes have different numbers of neighbors~\cite{hamilton2020graph}. GNNs have instead been conceived to effectively operate on graph-structured data~\cite{wu2020comprehensive}. 
% Precisely, GNNs generate representations of nodes i.e., node embeddings that depend on the structure of the graph and the feature information of the nodes. 

GNNs implement complex functions that are capable of encoding the properties of a graph's nodes in terms of low-dimensional vectors, called embeddings, which summarize their position within the graph, their  features, and the structure and properties of their local neighborhood. In other words, GNNs project nodes into a latent space such that a similarity in the latent space approximates a similarity in the original graph~\cite{hamilton2020graph}. To do this, GNNs use a form of neural message-passing algorithm, where vector messages are exchanged between nodes and updated using neural networks~\cite{gilmer2017neural}. We provide more details about GNNs by first recalling the definition of a graph. 
\begin{definition}[Graph]
A graph $\mathcal{G} = \mathcal{(V, E)}$ is defined by a set of nodes $\mathcal{V}$ and a set of edges $\mathcal{E}$ between these nodes. An edge going from node \(u \in\) $\mathcal{V}$ to a node \(v \in\) $\mathcal{V}$ is denoted as \((u, v) \in\) $\mathcal{E}$. The neighborhood of a node \(v\) is given by $\mathcal{N}(v) = \{u \in \mathcal{V} : (u,v) \in \mathcal{E}\}$. The \emph{attribute} or \emph{feature} information associated with  node $v$ is denoted by \(x_v \in \mathbb{R}^n\). $ \mathbf{X} \in \mathbb{R}^{|\mathcal{V}|\times n}$ represents the node feature matrix of the graph. 
\end{definition}
%Similarly, let \(x_{u,v}^e\in \mathbb{R}^c\) be the feature vector of an edge between \({u,v}\) and \(\textbf{X}^e \in \mathbb{R}^{m\times c}\) represents the edge feature matrix of the graph. 
%The success of deep learning in different domains is partially attributed to the powerful computational resource availability, availability of big training data and the ability of deep learning to effectively extract latent representations from Euclidean data like images, and texts.
%CNNs can learn hidden patterns of Euclidean data effectively but many naturally occurring data is often not in Euclidean form, graph being an example. The complexity of graph data structure imposes significant challenges on existing machine learning algorithms since graphs are irregular, has variable number of nodes, and different nodes have different number of neighbors implying that important operations like convolution may be difficult to apply to the graph domain as is. 
%Furthermore, a core assumption of existing machine learning algorithms that each instance is independent of each other no longer holds true for graph data because each instance i.e., a node is related to its neighbors. 
%This give rise to the need for new functions that can deal with graph data structures. Infact, Graph Neural Networks (GNNs)~\cite{gori2005new, scarselli2008graph} can deals with graphs and Z. Wu et al. provides a comprehensive survey on GNNs in~\cite{wu2020comprehensive}. 

During the message-passing step of the $k^{th}$ GNN layer, an embedding $h_u^{(k)}$ corresponding to each node $u \in \mathcal{V}$ is updated according to the information aggregated from its graph neighborhood $\mathcal{N}(u)$. We can express this message-passing update as:
\begin{equation}
    m_{\mathcal{N}(u)}^{(k)} = \textsc{Aggregate}^{(k)}(\{h_v^{(k-1)}, v \in N(u)\}),
\end{equation}
\begin{equation}
h_u^{(k)} = \textsc{Update}^{(k)}(
%\{
{h_u}^{(k-1)}, m_{\mathcal{N}(u)}^{(k)}
% \}
), 
\end{equation}
where \textsc{Aggregate} and \textsc{Update} are differentiable functions and $m_{\mathcal{N}(u)}^{(k)}$ is the ``message'' aggregated from the graph neighborhood $\mathcal{N}(u)$ of $u$. At each GNN layer, %\pierluigi{what does it mean an iteration of a neural network? Do you mean layer? Or maybe iteration of the training algorithm? Or maybe inference?} 
the \textsc{Aggregate} function takes as input the set of embeddings of the nodes in the neighborhood $\mathcal{N}(u)$ of $u$  and generates a message $m_{\mathcal{N}(u)}^{(k)}$ based on this aggregated information. The \textsc{Update} function then combines the message $m_{\mathcal{N}(u)}^{(k)}$ with the previous embedding $h_u^{(k-1)}$ of node $u$ to generate the updated embedding $h_u^{(k)}$. In this way, GNNs encode the node feature information and the aggregated neighborhood information into the node embeddings.

The initial embeddings, at $k$ = 0, are the node features $x_v$. Since the \textsc{Aggregate} function takes a set as input so it should be permutation invariant in nature~\cite{hamilton2020graph}.
Once the node embeddings are generated, they can then be used to successfully perform different tasks like node classification~\cite{kipf2016semi, hamilton2017inductive}, link prediction~\cite{zhang2018link}, graph classification~\cite{zhang2018end, ying2018hierarchical} to name a few, in an end-to-end manner. %\pierluigi{Unclear. Are GNN machine learning systems too?} 
In the case of graph-related tasks like node classification, GNNs are trained to generate node embeddings such that the embeddings of the nodes of a given class are close to each other in the latent space. In this work, we leverage GNNs to formalize the problem of classifying the registers in a netlists as a graph node classification problem. 

\begin{figure}[t]
  \centering
  \includegraphics[width=\columnwidth]{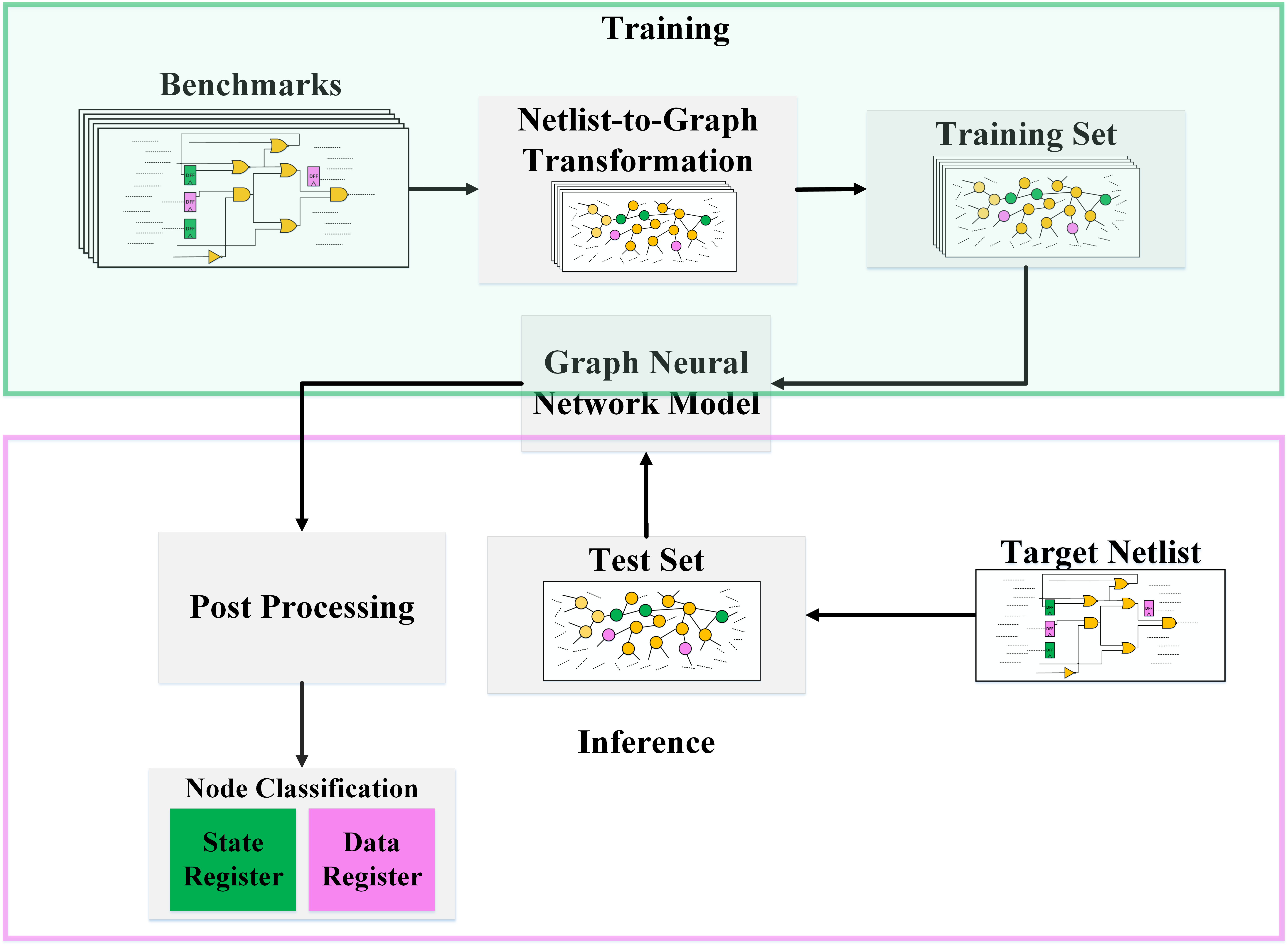}
  \caption{ReIGNN methodology.}
  \label{fig:reignn_flow}
   \vspace{-4mm}
\end{figure}

\section {ReIGNN: State Register Identification Using Graph Neural Networks} \label{sec:reginn}

Figure~\ref{fig:reignn_flow} illustrates the ReIGNN framework for classifying state and data registers in a netlist. We first train the GNN pipeline with a set of benchmarks and then leverage the trained GNN to perform inference on new netlists. The first step in training is to transform a circuit netlist into a graph $\mathcal{G}$ and associate the graph nodes with feature vectors and labels as described in Section~\ref{sec:reignn_feature}. In the second step, the graphs are passed to the GNN framework to generate embeddings for the register nodes.  Finally, the embeddings are processed by a multi-layer perceptron (MLP) stage with softmax activation functions to perform inference, i.e., compute the predicted labels for the register nodes in the design, as described in Section~\ref{sec:reginn_gnn}. During the training phase, several benchmarks are used for learning the weight parameters of the GNN layers and the MLP layer. %\pierluigi{Is there a way to decide how many? 1 million?} 
% Once the training is complete, the GNN pipeline is used for classifying registers of new unseen circuit netlists. 
After GNN inference, we further improve the classification accuracy by performing structural analysis. We detail these steps below. 

\begin{figure*}[t]
  \centering
  \begin{tabular}{cc}
 \includegraphics[width=\columnwidth]{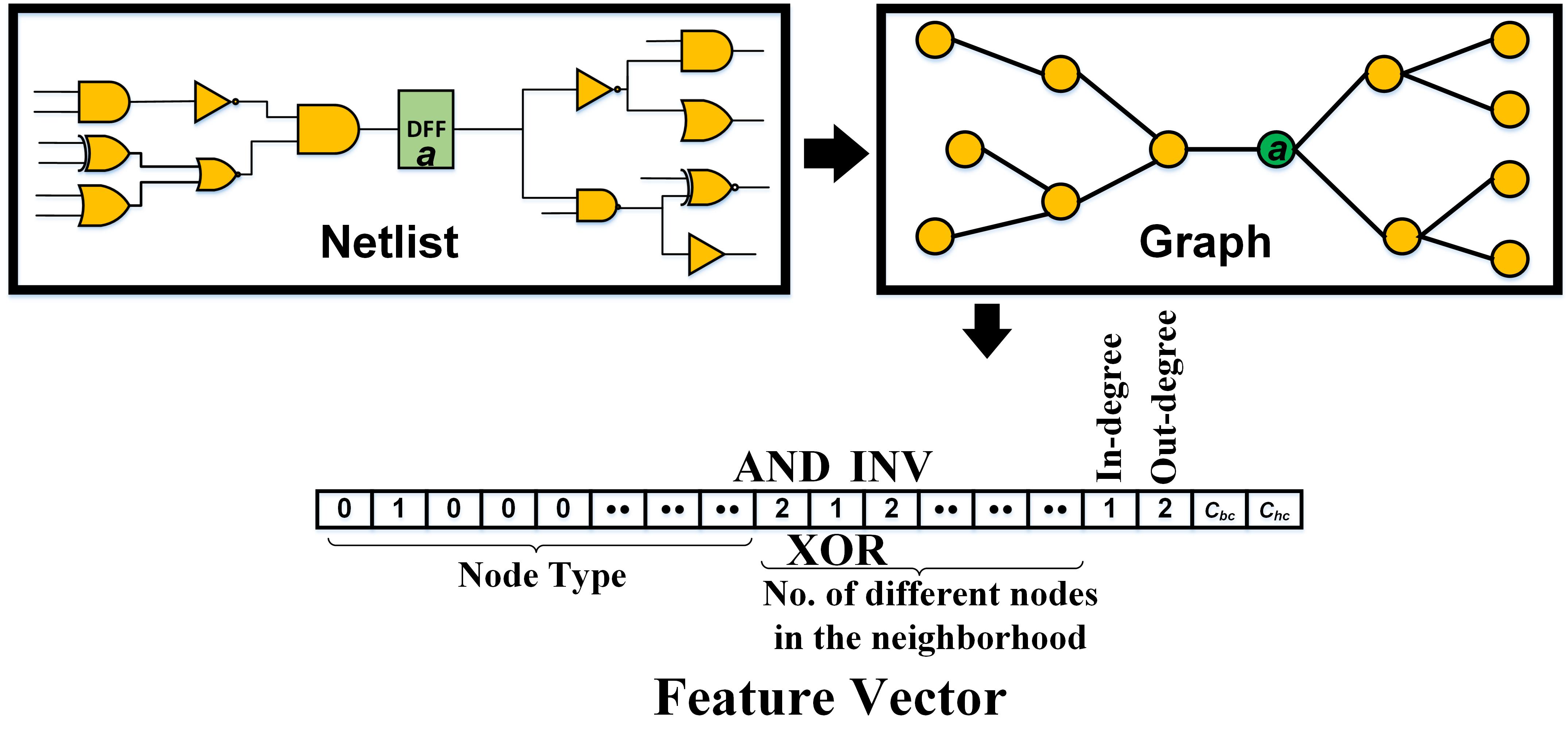} &   \includegraphics[width=\columnwidth]{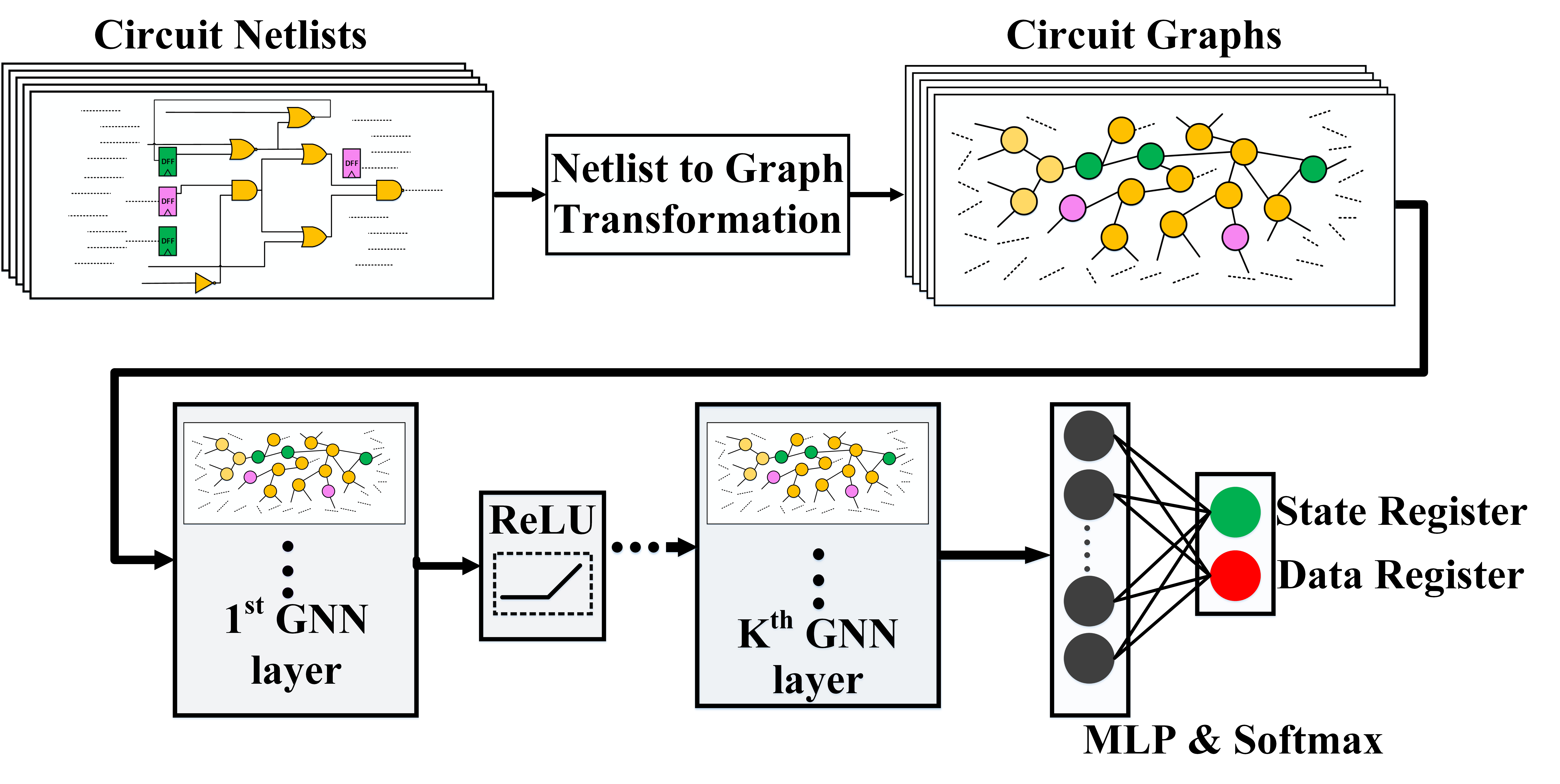}\\
       (a)  & (b)
  \end{tabular}
  \caption{ReIGNN methodology workflow (a) Feature vector extraction and representation for the node \(a\) in the netlist sub-graph, (b) GNN model architecture.
  }
  \label{fig:graph_gnn}
  \vspace{-4mm}
\end{figure*}

\subsection {
% Netlist to graph transformation and 
Node Feature Identification and Extraction} \label{sec:reignn_feature}

% In order to perform learning tasks on the gate-level netlists, 
We translate a gate-level netlist into a generic, technology-independent representation in terms of  a graph $\mathcal{G}$, where the gates are the nodes and the wires connecting the gates become the edges.  
%is connectivity between the gates in the circuit netlist using a graph \(G = (V, E)\) where the set \(V\) of nodes represent all the gates and registers in the netlist while the set \(E\) of edges represent the wires. 
Unlike previous register identification techniques~\cite{meade2016gate, brunner2019improving}, 
ReIGNN does not require processing the netlist to only include AND, OR, and INV cells. 
% is directly converted into a graph. 
An example of a circuit netlist and its corresponding graph is shown in Fig.~\ref{fig:graph_gnn}a. 

Each node in the graph is then associated with a feature vector by using functional properties of the logic cells and node-level statistics of the graph. The feature vector $x$ of a node contains the following information: (a) node type, (b) node in-degree, (c) node out-degree, (d) node betweenness centrality, (e) node harmonic centrality, and (f) number of different types of gates appearing in the neighborhood of the node. We use one-hot encoding to represent the node type. For example, if a target library has in total $10$ different types of logic cells, then we have a $10$-dimensional vector representing the node type. Changes in the target library may then impact $|x|$. The \emph{node in-degree} is the number of incoming neighbors, while the \emph{node out-degree} is the number of outgoing neighbors. The \emph{betweenness centrality} and \emph{harmonic centrality} of a node in a graph are defined as follows.
\begin{definition}[Betweenness Centrality~\cite{brandes2008variants}]
Betweenness centrality of a node \(v\) is the sum of the fraction of all-pairs shortest paths that pass through \(v\), i.e., 
\begin{equation}
{C_{bc}}(v) = \sum_{x,y\in \mathcal{V}}\frac{\sigma(x,y)|v}{\sigma(x,y)}
\end{equation}
where \(\sigma(x,y)\) is the number of shortest paths between nodes \(x\) and \(y\) and \(\sigma(x,y)|v\) is the number of those paths passing through the node \(v\), with \(v \ne x,y\).
\end{definition}

\begin{definition}[Harmonic Centrality~\cite{boldi2014axioms}]
Harmonic centrality of a node \(v\) is the sum of the reciprocal of the shortest path distances from all other nodes to \(v\), i.e., 
\begin{equation}
{C_{hc}}(v) = \sum_{u \ne v}\frac{1}{d(u,v)}
\end{equation}
where \(d(u,v)\) is the shortest distance between \(u\) and \(v\).
\end{definition}

% \textcolor{blue}{
Intuitively, the betweenness centrality quantifies the number of times a node acts as a bridge along the shortest paths between any two other nodes. Nodes with high betweenness centrality usually have more influence over the information flowing between other nodes. Since the control logic in a design is in charge of sending signals to control the data-path blocks, the betweenness centrality is expected to be higher for the state registers than the data registers. Computing the betweenness centrality is quadratic in the number of nodes in a graph. However, we can reduce the computational cost by resorting to an approximate betweenness centrality metric~\cite{brandes2008variants}.
% }

% \textcolor{blue}{
On the other hand, the harmonic centrality is inversely proportional to the average distance of a node from all other nodes. Because the control logic sends control signals across the design, harmonic centrality is also expected to be higher for state registers.
% }
%\pierluigi{Any clue on why all these make sense to be used? Can we skip one of them and maybe it works the same?}
% We use all of the above information to create the feature vector for each node in the graph. 
Figure~\ref{fig:graph_gnn} shows an example feature vector associated with the node $a$ in the graph. For the training and validation dataset, we also have labels $y$ associated with the nodes in the graph. Hence, the \emph{training and validation set} is a set of graphs such that each node is provided with a feature vector and a label.
%Let us consider an example ---- Explain this in the experimental section of the paper. 
%and Table 1 shows the list of these gates and the encoding we use for indicating the gate type. Here, we use 1-hot encoding and the values in the table are written in decimal format to show the table~\ref{table:1} in a compact format. 

\subsection {Graph Learning Pipeline} \label{sec:reginn_gnn}

Once the training set is created, the next step is to train the graph learning pipeline. The ReIGNN framework processes the graph data and generates node embeddings, as illustrated in Fig.~\ref{fig:graph_gnn}b. We use GraphSAGE~\cite{hamilton2017inductive} as the GNN layer,  
% Many of the existing GNN techniques~\cite{wu2020comprehensive, kipf2016semi}  are inherently transductive. They train individual embeddings for each node in the graph using matrix-factorization-based objectives, and do not naturally generalize to unseen data i.e., new graphs 
%since they make predictions on nodes in a single, fixed graph
% ~\cite{hamilton2017inductive}. On the other hand, GraphSAGE, 
which follows an inductive approach to generate node embeddings to facilitate generalization across unseen graphs. 
%GraphSAGE leverages node features to learn aggregator functions that can generate embeddings for unseen nodes and graphs and thus facilitate generalization across graphs. 
Instead of training a distinct embedding vector for each node, GraphSAGE trains an aggregator function that learns to aggregate feature information from a node's local neighborhood to generate node embeddings. Therefore, GraphSAGE simultaneously learns the topological structure of each node’s neighborhood and the distribution of node features in the neighborhood~\cite{hamilton2017inductive}.
% GraphSAGE, because of its inductive approach, is more appropriate for our application since we need to classify the registers of new unseen netlists. 
%\pierluigi{This paragraph is not clear (what is an inductive approach and why it is good?). All the wording should be improved to be precise.}

We use the mean aggregation function for the GraphSAGE layers to update the embedding for each node~\cite{hamilton2017inductive}. In each iteration $k$ of message propagation, the GraphSAGE layer updates the node embedding \({h_v}^{k}\) of node \(v\) as
\begin{equation}
%{\textbf{h}_v}^{k} = \sigma({\textbf{W}_1}\cdot\ {\textbf{h}_v}^{k-1}+\textbf{W}_2 \cdot MEAN({h_u}^{k-1}, \forall u \in N(v)))
h_v^{k} \leftarrow \sigma(\textbf{W}^k\cdot \textsc{Mean}(\{h_v^{k-1}\}\cup \{{h_u}^{k-1}, \forall u \in N(v)\})),
\label{eq:sageconv}
\end{equation}
where $\textbf{W}^k$ is a trainable weight parameter matrix associated with the $k^{th}$ GraphSAGE layer, \textsc{Mean} is the aggregator function, and $h_v^{k}$ represents the embedding for node $v$ at the $k^{th}$ layer. The \textsc{Aggregate} and \textsc{Update} functions are then combined together in~\eqref{eq:sageconv}. For the first layer, $h_v^{0}$ is the node feature vector $x_v$. $\sigma(\cdot)$ is the activation function, i.e., a rectified linear unit (ReLU). We also denote by \(h_v^{fin}\) the final node embedding. Following the final GNN layer, there is an MLP layer with \textsc{Softmax} activation function, which processes the node embeddings \(h_v^{fin}\) to produce the final prediction $ z =  $\textsc{Softmax}$(\textsc{MLP}(h_v^{fin}))$ providing a measure of the likelihood of the two classes, i.e., `state' and  `data' registers. The class with higher likelihood will be taken as the result. 

A circuit graph contains both logic gates and registers. While all nodes and their incident edges in the graph are considered in the message passing operations to generate the embeddings, only the final layer embeddings $h_v^{fin}$ of the register nodes are used to compute the loss function. To train the model, we use the negative log-likelihood loss. %\pierluigi{Unclear. Not well explained.}
%\begin{equation}
%\mathcal{L} = \sum_{u %\in \mathcal{V}_{train}} - \log(\textsc{softmax}({z}_u,{y}_u))
%\label{eq:lossfn}
%\end{equation}

%where $y_u$ is the true label, $z_u$ is the \textsc{softmax} function output of the node $u$.  %\mathcal{V}_{train}$ represents the set of register nodes which are used for calculating the loss. 
%So we consider all the nodes and their incident edges in the message passing operations and the node embedding $h_{u}^{k}$ is generated for all the nodes but while computing the loss, we consider only the registers in the graph. 

\subsection {Post Processing} \label{sec:reignn_post_pc}

We leverage structural information associated with the gate-level netlist graphs to further improve the classification accuracy of ReIGNN. 
As shown in Fig.~\ref{fig:fsm}, the state registers in the control logic of a design must have a feedback path and they should all be part of a strongly connected component (SCC)~\cite{fyrbiak2018difficulty}, defined as follows.

\begin{definition}[Strongly Connected Component]
A strongly connected component of a directed graph $\mathcal{G}$ is a maximal subset of vertices in $\mathcal{V}$ such that for every pair of vertices $u$ and $v$, there is a directed path from $u$ to $v$ and a directed path from $v$ to $u$. \end{definition}

An SCC in a graph can be obtained using Tarjan's algorithm~\cite{tarjan1972depth}. %\pierluigi{Is that unique?}. 
After the GNN prediction, we use the circuit connectivity information to rectify any misclassifications of state registers. In particular, we check if a register identified by the GNN as a state register is part of an SCC and has a feedback loop. If any of these criteria is not satisfied, we safely reclassify the register as a data register. 

\section{Evaluation} \label{sec:eval}

We provide details about our dataset and  evaluation metrics, and report the performance of ReIGNN on different benchmarks by comparing it with  RELIC~\cite{meade2016gate}. Our framework % Netlist to graph transformation and efficient node feature extraction 
is implemented in Python and we use  
% using the NetworkX~\cite{hagberg2008exploring} package. We have used 
the Pytorch Geometric package~\cite{Fey/Lenssen/2019} for implementing the graph learning model. We implemented RELIC by following Algorithm~\ref{alg:relic}. Training and testing of the model were performed on a server with 
% \pierluigi{I think more than the graphics card what matters are the GPUs and how many? Don't use anything commercial in a paper. You are not doing advertisement.} % 
an NVIDIA GeForce RTX 2080 graphics card, 48 2.1-GHz processor cores, and 500-GB memory. 

\begin{table}[t]
    \centering
    \caption{Overview of the Benchmarks}
    \resizebox{\columnwidth}{!}{
    {\makegapedcells
    \begin{tabular}{c c c c c} 
    \hline
    \textbf{Design Name} & \textbf{Total No. Inputs} & \textbf{Total No. Registers} &  \textbf{No. State Registers} & \textbf{Total No. Gates}\\
    \hline
    $\textup{aes}^{\star}$ & 45 & 2994 & 15 & 29037\\
    \hline
    $\textup{siphash}^{\star}$ & 44 & 794 & 8 & 6214\\
    \hline
    $\textup{sha1}^{\star}$ & 516 & 1526 & 3 & 11822\\
    \hline
    $\textup{fsm}^{\diamond}$ & 17 & 15 & 7 & 166\\
    \hline
    $\textup{gpio}^{\diamond}$ & 111 & 51 & 11 & 311\\
    \hline
    $\textup{memory}^{\diamond}$ & 173 & 75 & 7 & 881\\
    \hline
    %\(b10^{\dagger}\) & 12 & 17 & 4 \\
    %\hline
    $\textup{uart}^{\wedge}$ & 12 & 69 & 10 & 469\\
    \hline
    %$\textsc{CPU8080}^{\wedge}$ & 12 & 275 & 38 \\
    %\hline
    $\textup{cr}\_\textup{div}^{\wedge}$ & 99 & 4172 & 4 & 34218\\
    \hline
    $\textup{altor32}\_\textup{lite}^{\wedge}$ & 39 & 1249 & 6 & 13111\\
    \hline
    $\textup{gcm}\_\textup{aes}^{\wedge}$ & 267 & 1697 & 10 & 34496\\
    \hline
    \end{tabular}}}
    \label{table:benchmarks}
        \vspace{-4mm}
\end{table}

\subsubsection{Dataset Creation} \label{5.1}

We assess the performance of ReIGNN on a set of standard benchmark circuits, which were also used in the literature~\cite{meade2016gate, brunner2019improving}. As summarized in  Table~\ref{table:benchmarks}, the benchmarks include designs from OpenCores \((\wedge)\)~\cite{opencores}, the secworks github repository  \((\star)\)~\cite{secworks}, and blocks from a $32$-bit RISC-V processor and a $32$-bit microcontroller \((\diamond)\)~\cite{onchipuis}. We synthesized these benchmark circuits using Synopsys Design Compiler with a 45-nm Nangate open cell library\cite{nangate}. %\pierluigi{references?}. 
% Since the control logic FSM of a design can use different types of encoding, 
We considered both a one-hot encoded and a binary encoded version of each design while creating the dataset. Table~\ref{table:benchmarks} shows the number of inputs, total number of registers, and the number of state registers for the one-hot encoded version of the designs. We get the number of inputs and total number of registers in a design from the synthesized netlist, while the number of state registers is obtained from the RTL description. 

Different synthesis constraints are used to synthesize different versions of the benchmark circuits in Table~\ref{table:benchmarks}. Specifically, each benchmark in the table is synthesized according to $4$ different constraint configurations, resulting in a total of $40$ different circuit designs, providing the one-hot encoded dataset. We use different synthesis constraints to achieve different neighborhood configurations for the nodes in the circuit. The binary encoded dataset, which also contains $40$ benchmarks, is generated in a similar way. ReIGNN transforms the netlists into graphs and associates each node in the graphs with a feature vector. 
% \textcolor{blue}{
The one-hot encoded dataset contains in total 618,602 nodes and 1,123,713 edges, while the binary encoded dataset has 616,001 nodes and 1,119,637 edges. We use the RTL description of the netlists to obtain the labels for the register nodes in the graphs. For the Nangate library, the size of the feature vector
%has in total 9 different types of basic gates, so for 1-hot encoding of the type of node in a graph we need a vector of size 11 since the node could be anyone of those 9 gates or a primary input or a flip-flop. Thus, 
${x}$ of each node is $26$, considering all the features mentioned in Section~\ref{sec:reignn_feature}. 
% Changing the target library only affects \(|\textbf{x}_v|\).
% } 
% \pierluigi{Unclear sentence needs rephrasing or explanations. What is this? Explanation? Reference?}\Subhajit{should we move this to the evaluation method and metrics section?}
%Each time the graphs associated with a particular benchmark creates the test set, while 80\% of the remaining graphs creates the training set and the other 20\% creates the validation set.}
%ReIGNN classifies the registers of each design separately by excluding its corresponding graphs from the training/validation set. For example, when trying to classify the registers of \textsc{AES}, all the different synthesized graphs of \textsc{AES} forms the test set while the graphs of the other benchmarks are divided into the training and validation set to avoid biasing.

\subsubsection{GNN Model Configuration} \label{sec:res_topology}

In the experiments, the architecture of ReIGNN contains $3$ GraphSAGE layers, each with $100$ hidden units, followed by the MLP layer with softmax activation units.
% \textcolor{blue}{
The number of GNN layers, determined empirically, is based on the size of the neighborhood under consideration for each gate. 
% We conducted experiments where we sweeped the number of GNN layers and found that 3 is the optimal number of layers.
% }
%During training, we add dropout layers with a rate of $0.25$ after each GraphSAGE and MLP layer. 
% to avoid over-fitting of the model and facilitate regularization.
% \textcolor{blue}{
To avoid overfitting of the trained model, we also use dropout, a regularization technique that randomly drops weights from the neural network during training~\cite{srivastava2014dropout}.
%} 
%\pierluigi{Unclear sentence needs rephrasing or explanations. Is ``dropout'' the name of a technique? What is the relationship between dropout and dropout layers? What are ``activations''?}
We then add normalization layers after each GraphSAGE layer. The details of the GNN model as well as the training architecture is provided in Table~\ref{table:2}. 
% \textcolor{blue}{
As the dataset is highly imbalanced (the number of data registers is much larger than the number of state registers), we use a weighted loss function inducing a greater penalty for misclassifying the minority class. The weights are set based on the ratio between state and data registers in the dataset.
%}
% \pierluigi{Are these things normal or legitimate? Unclear.}
%As done in Sec.~\ref{sec:res_topology}

\begin{table}[t]
    \centering
    \caption{GNN Configuration and Training Details}
    {\makegapedcells
    \begin{tabular}{c c c c} 
    \hline
    & \textbf{Architecture} &  
     \textbf{Training Settings} & \\
    \hline
    {Input Layer} & \([|{x}|, 100]\) & {Optimizer} & Adam \\
    \hline
    Hidden Layer & \([100, 100]\) & {Learning Rate} & 0.001 \\
    \hline
    MLP Layer & \([100, 50]\)  &{Dropout} & 0.25 \\
    \hline
    Output Layer & [50, \#classes] &{No. of Epochs} & 300 \\
    \hline
    {Activation} & ReLU \\
    \hline
    {Classification} & Softmax \\
    \hline
    \end{tabular}}
    \label{table:2}
    \vspace{-3mm}
\end{table}

\begin{figure*}[t]
  \centering
  \includegraphics[width=2.0\columnwidth]{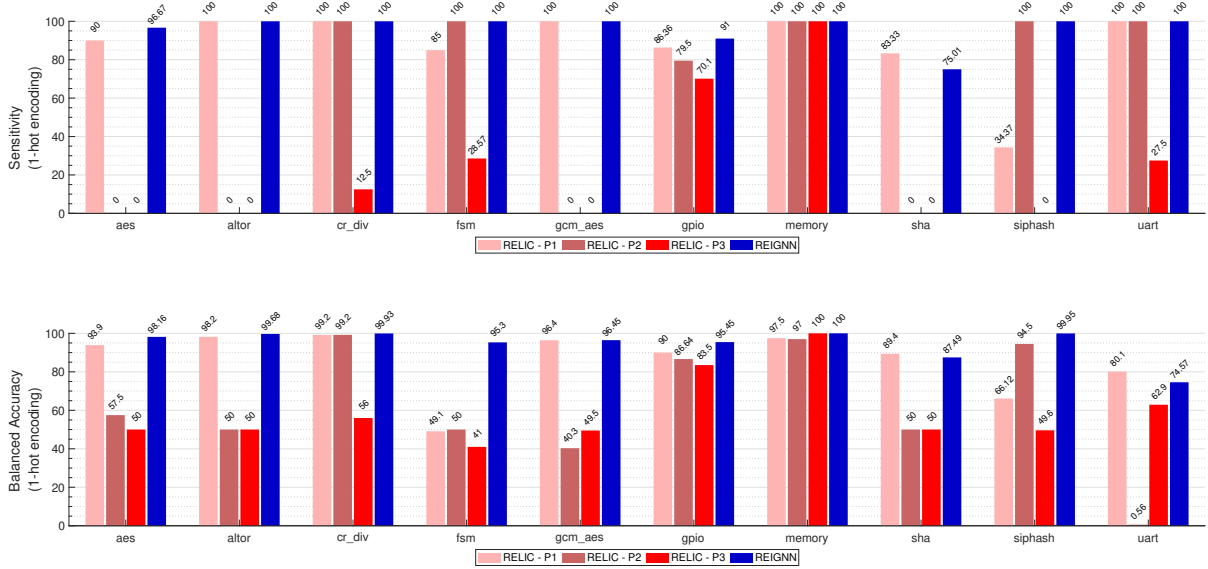}
  \caption{Performance comparison between ReIGNN and RELIC on one-hot encoded benchmarks.}
  \vspace{-6mm}
  \label{figure_experiment_1hot_encoding}
\end{figure*}

\subsubsection{Evaluation Method and Metrics} \label{5.3}

We perform classification tasks on an independent set of netlists, which excludes the ones used for training and validation. Specifically, we use \emph{k}-fold cross validation~\cite{kohavi1995study}, where the data is initially divided into \emph{k} bins and the model is trained using \emph{k}-1 bins and tested on the \emph{k}-th bin. This process is repeated \emph{k} times and each time a new bin is considered as the test set. In our case, the $4$ different versions of a design create the test set each time. For example, when reporting the performance on the  $aes$ benchmark in the one-hot encoded dataset, $4$ different $aes$ graphs create the test set, $80\%$ of the remaining graphs in the dataset are used for training, while the rest are used for validation. 
% Once we have the predictions from the trained model, ReIGNN performs post-processing and outputs the final node classification result. 
% \textcolor{blue}{
To evaluate the performance of ReIGNN, we compare its  predictions with the true labels of the nodes and use the standard statistical measures of correctness of a binary classification test. Since the dataset is imbalanced, we use sensitivity, or true positive rate (TPR), specificity, or true negative rate (TNR), and balanced accuracy as the metrics, defined as follows:
%} 
%
\begin{equation}
%\textbf{Sensitivity} = \frac{\text{No of true positives}}{\text{No of true positives + No of false Negatives}}
\text{Sensitivity} = \frac{\text{No. true positives}}{\text{No. true positives + No. false negatives}},
\end{equation}
\begin{equation}
%\textbf{Sensitivity} = \frac{\text{No of true positives}}{\text{No of true positives + No of false Negatives}}
\text{Specificity} = \frac{\text{No. true negatives}}{\text{No. true negatives + No. false positives}},
\end{equation}
\begin{equation}
%\textbf{Accuracy} = \frac{\text{No of true positives + No of True Negatives}}{\text{No of true positives + No of true negatives + No of false Positives + No of false Negatives}}
\text{Balanced Accuracy} = \frac{\text{Sensitivity + Specificity}}{\text{2}},
\end{equation}
where we consider detecting state registers as positive outcomes and data registers as negative outcomes. Therefore, a \emph{true positive} (TP) implies that a state register is correctly identified as a state register by ReIGNN, while a \emph{false positive} (FP) means that a data register is incorrectly classified by ReIGNN as a state register. \emph{True negatives} (TNs) and \emph{false negatives} (FNs) can also be defined in a similar way. Overall, the \emph{sensitivity} is the ratio between the number of correctly identified state registers and the total number of state registers in the design. The 
% \textcolor{blue}{
\emph{specificity} is the ratio between the number of correctly identified data registers and the total number of data registers. The \emph{balanced accuracy} is the average of sensitivity and specificity.
% }
% A sensitivity of $100\%$ implies that all the state registers were identified correctly. An accuracy of $100\%$ implies that all the state and data registers were identified correctly. 
High sensitivity and high balanced accuracy are both desirable, since they impact any FSM extraction methodology for IC reverse engineering. If we report low sensitivity, then there exist unidentified state registers, which will impact the correctness of the extracted FSM. On the other hand, low specificity, hence low balanced accuracy, makes FSM extraction time consuming since many data registers are also identified as state registers and will be included in the STG of the FSM. 
% \pierluigi{Some of this consideration should probably be anticipated to emphasize the impact and motivation of this work.}

\subsubsection{Performance Analysis}

%\textcolor{blue}{
We evaluate the performance of ReIGNN on all the  benchmarks from the one-hot encoded and binary encoded datasets. The GNN models are trained offline using the available data, thus incurring a one-time-only training cost. The average test time of the trained model is $1$~s while the average training time is $8500$~s.
% } 
For each benchmark,  Fig.~\ref{figure_experiment_1hot_encoding} shows the average balanced accuracy and sensitivity across the $4$ different synthesized versions of the netlist. We compare the achieved performance with the one of RELIC~\cite{meade2016gate}, which has a similar classification accuracy as fastRELIC~\cite{brunner2019improving}. In fact, while  fastRELIC reduces the run-time of RELIC, it uses the same heuristic for determining the similarity scores between two registers. 
% fan-in structure is same for both of them implying that their  is also similar for them. 
In RELIC, the user needs to set the three threshold values $T1$, $T2$, and $T3$ together with the fan-in cone depth \emph{d}. 
% \pierluigi{The role of these thresholds is still mysterious.}
%and there is no optimal parameter configuration that works for all designs and different parameter configuration impacts the outcome differently in each design. 
In our experiments, we consider the following parameter configurations, which are also used in the literature:
\begin{itemize}
    \item P1: $T1$ = 0.5, $T2$ = 0.8, $T3$ = 1, \emph{d} = 5; 
    \item P2: $T1$ = 0.7, $T2$ = 0.5, $T3$ = 5, \emph{d} = 5; 
    \item P3: $T1$ = 0.4, $T2$ = 0.5, $T3$ = 4, \emph{d} = 7.
\end{itemize}
% \pierluigi{Why these configurations make sense? Why not trying others?}
Depending on the threshold values, both balanced accuracy and sensitivity vary largely across different benchmarks. Across the one-hot encoded benchmarks, ReIGNN reports an average balanced classification accuracy of $94.7\%$ and a sensitivity of $96.3\%$. RELIC's average balanced accuracy varies between $68.14\%$ (for configuration P2) and $86.6\%$ (for configuration P1) while its sensitivity varies between $23.91\%$ (for configuration P3) and $87.96\%$ (for configuration P1).
% configuration P1 has an average classification accuracy of xx\% and a sensitivity of xx\%, for  parameter configuration P2 has an average classification accuracy of xx\% and a sensitivity of xx\%, and for  parameter configuration P3 has an average classification accuracy of xx\% and a sensitivity of xx\%. 
%%% Add that the sensitivity is 100% for most of the designs.

\begin{figure*}[t]
  \centering
  \includegraphics[width=2.0\columnwidth]{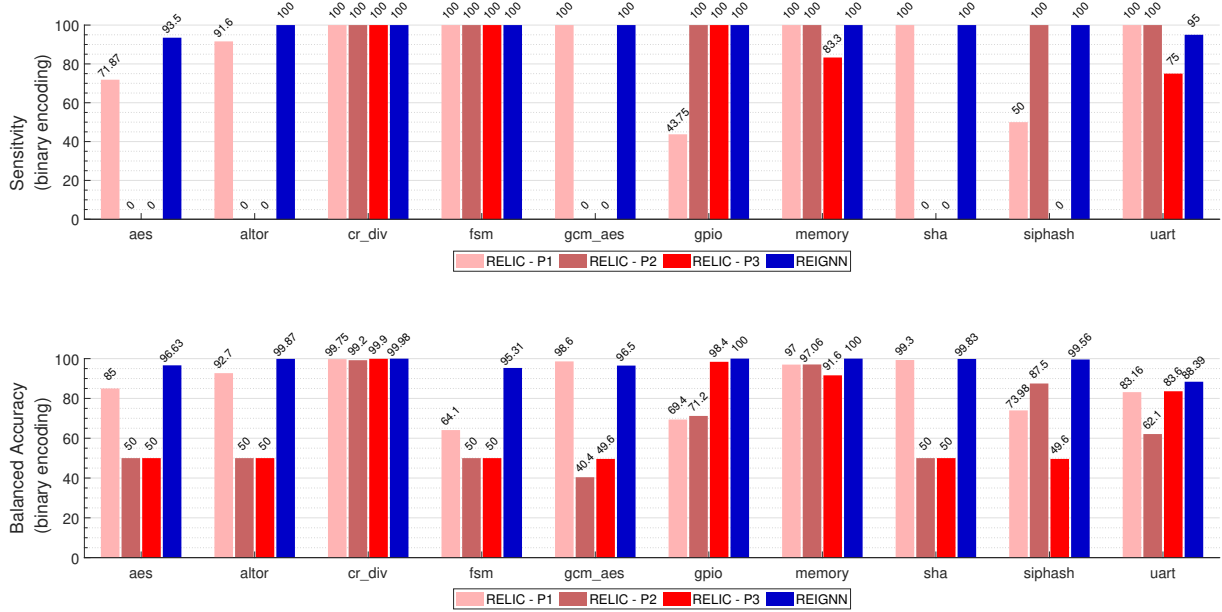}
  \caption{Performance comparison between ReIGNN and RELIC on binary encoded benchmarks.}
  \label{figure_experiment_binary_encoding}
  \vspace{-6mm}
\end{figure*}

To analyze the impact of the performance of different tools, we consider, for instance, the $aes$ benchmark, including a total of $2994$ registers, only $15$ of which are state registers. 
% while 2979 are data registers.
As shown in Fig.~\ref{figure_experiment_1hot_encoding}, across the four synthesized versions of the netlist, ReIGNN achieves on average a balanced accuracy of $98.16\%$ (FP = $11, 11, 10, 9$  for the four netlists) and a sensitivity of $96.67\%$. Specifically, the sensitivity is $100\%$ (TP = $15$, FN = $0$) for two designs and decreases to $93.33\%$ (TP = $14$, FN = $1$) for the other two. On the other hand, when using configuration P1, the average accuracy of RELIC is $93.9\%$ with FP values of $71$, $67$, $55$, and $69$. RELIC achieved a sensitivity of $80\%$ for two netlists (TP = $12$, FN = $3$) and $93.33\%$ (TP = $14$, FN = $1$) and $100\%$ (TP = $15$, FN = $0$) on the other two. Overall, Algorithm~\ref{alg:relic} produces more FPs, which can make the FSM extraction task harder. 
% since now many data registers will be considered in the analysis since they were wrongly classified as state registers by RELIC. 
% We recall that in case of an FSM, all the state registers are a part of an SCC. This SCC can contain data registers too.

Whenever ReIGNN cannot achieve full sensitivity, meaning that some state registers are misclassified as data registers, we can still use structural analysis to bring the number of FNs to zero. In fact, by finding the SCCs associated with the correctly classified state registers, we can look for all the registers that are part of the SCCs, some of which may indeed be data registers, and label them as state registers for FSM extraction. This extension makes ReIGNN complete (TN = $0$) but compromises its soundness (FP $\neq 0$), which, again, may increase the complexity of the FSM extraction. Finally, when using configuration P2 and P3, Algorithm~\ref{alg:relic} has $0\%$ sensitivity, meaning that all state registers are misclassified and it is not possible to recover the FSM. 
% 
% \textcolor{blue}{
As shown in Fig.~\ref{figure_experiment_1hot_encoding}, ReIGNN can achieve an average sensitivity of $100\%$ for uart but the average balanced accuracy is only  $74.57\%$, implying that the number of FPs is high. 
% The trained GNN misclassifies many data registers as state registers and structural analysis cannot help rectify this misclassifications. 
This is due to the fact that 
% Upon further analysis of the UART benchmark, we observe that 
the registers associated with the counters in the design are all classified as state registers by the GNN. Moreover, since these registers are all part of an SCC, structural analysis cannot help rectify these misclassifications. While a counter is indeed an FSM, by our definition, its registers are data registers, in that they do not hold the state bits of the control logic FSM. This explains the lower performance of ReIGNN. 
% Hence in this case, the registers of the counters are considered as data register since they are not part of the control logic of the design.}
% The accuracy for configuration P2 and P3 are high but that has or no impact since the sensitivity is 0\%.

Across the binary encoded benchmarks, ReIGNN reports an average balanced accuracy of $97.9\%$ and a sensitivity of $99.4\%$. RELIC's average balanced accuracy varies between $66.4\%$ (for configuration P2) and $86.6\%$ (for configuration P1) while its sensitivity varies between $45.8\%$ (for configuration P3) and $85.72\%$ (for configuration P1)
as shown in Fig.~\ref{figure_experiment_binary_encoding}. 
% \textcolor{blue}{
In general, structural analysis has a considerable impact on the overall performance of ReIGNN. While the sensitivity remains unchanged, as explained in Section~\ref{sec:reignn_post_pc}, the average balanced accuracy can substantially improve, e.g., by $6.52\%$, as shown in Table~\ref{table:impact of scc}, for the one-hot encoded benchmarks. While the application of structural analysis does not bring large improvements for designs like uart and gcm\_aes, due to the presence of the counters, 
% hence structural analysis cannot help reduce the FPs since the registers in the counters remain in an SCC. In other words, the specificity do not show
the balanced accuracy improves by $32.3\%$, $12.15\%$, and $11.3\%$ for designs like fsm, gpio, and memory, respectively.
% } % \pierluigi{Can we say here what we learn from the table? Can we quantify to help the reader interpret the table?}

\begin{table}[t]   
    \centering
    \caption{Analysis of the balanced accuracy change in ReIGNN due to inclusion of structural analysis}
	\resizebox{\columnwidth}{!}{
	{\makegapedcells
    \begin{tabular}{c c c}
    \hline
    \textbf{Design} & \textbf{Balanced Accuracy(GNN only)} & \textbf{Balanced Accuracy(GNN \& SCC)} \\
    \hline
    aes & 97.05\% & 98.16\% \\
    \hline
    altor32\_lite & 98.43\% & 99.68\% \\
    \hline
    cr\_div & 98.5\% & 99.93\% \\
    \hline
    fsm & 63\% & 95.3\% \\
    \hline
    gcm\_aes & 91.3\% & 96.45\% \\
    \hline
    gpio & 83.3\% & 95.45\% \\
    \hline
    memory & 88.3\% & 100\% \\
    \hline
    sha & 84.2\% & 87.49\% \\
    \hline
    siphash & 94.45\% & 99.95\% \\
    \hline
    uart & 73.3\% & 74.57\% \\
    \hline
    \end{tabular}}}
    \label{table:impact of scc}
    \vspace{-4mm}
\end{table}

Finally, to see how well ReIGNN generalizes across different types of benchmarks and encoding styles, we create a dataset with both one-hot encoded and  binary-encoded versions of each benchmark, for a total of $80$ graphs. Similarly to our previous experiments, we perform classification tasks on each benchmark independently. As shown in Table~\ref{table:experiment_combined},  
% summarizes the performance of ReIGNN in classifying the registers for different benchmarks. This table shows that 
ReIGNN generalizes well across different encoding styles and achieves on average a balanced classification accuracy of $96.93\%$ and a sensitivity of $97.53\%$.

\begin{table}[t]   
    \centering
    \caption{Performance of ReIGNN for combined dataset}
    \setlength{\tabcolsep}{7mm}{
    \begin{tabular}{ccc}
    \hline
    \textbf{Design} & \textbf{Sensitivity} & \textbf{Balanced Accuracy} \\
    \hline
    aes & 100\% & 99.69\% \\
    \hline
    altor32\_lite & 100\% & 99.49\% \\
    \hline
    cr\_div & 100\% & 99.99\% \\
    \hline
    fsm & 100\% & 93.75\% \\
    \hline
    gcm\_aes & 100\% & 96.76\% \\
    \hline
    gpio & 92.04\% & 96.02\% \\
    \hline
    memory & 100\% & 100\% \\
    \hline
    sha & 89.58\% & 99.94\% \\
    \hline
    siphash & 100\% & 99.9\% \\
    \hline
    uart & 93.75\% & 83.74\% \\
    \hline
    \end{tabular}}
    \vspace{-4mm}
    \label{table:experiment_combined}
\end{table}

%First we create 2 datasets namely Dataset 1 and Dataset 2. Dataset 1 consists of 1-hot encoded circuit netlists only and it has in total 55 different netlists. Dataset 2 consists of binary encoded version of the same circuits as dataset 1. As mentioned earlier, test set consist of the benchmark for which we measure the sensitivity and accuracy, while 60\% of the rest of the benchmarks are used for training and the rest 40\% creates the validation set. 

\section{Conclusions} \label{sec:conclusions}

We presented ReIGNN, a learning-based register classification methodology for finite state machine extraction and circuit reverse engineering. ReIGNN combines graph neural networks with structural analysis to classify the registers in a circuit with high accuracy and generalize well across different designs. Numerical results show the advantage of combining graph-based deep learning with post-processing based on structural analysis to achieve higher balanced accuracy and sensitivity with respect to previous classification methods. 
% \textcolor{blue}{
Future work includes further investigation of the proposed architecture to possibly reduce its complexity and the number of features for the same performance. 
\vspace{-1mm}
% ablation study on the current features to see if we can reduce the number of features yet maintain the similar level of sensitivity and balanced accuracy scores.}  

% \textcolor{blue}{
\section*{Acknowledgments}
This work was supported in part by the Air Force Research Laboratory (AFRL) and the Defense Advanced Research Projects Agency (DARPA) under agreement number FA8650-18-1-7817.
% }
\vspace{-1mm}
\bibliographystyle{ieeetr} 
%\bibliography{references}
\bibliography{conference_101719.bbl}

\end{document}